\documentclass[12pt]{article}
\usepackage{amssymb,amsmath,epsfig}

\begin{document}

\title{\bf Consistency of Anisotropic Inflation during Rapid Oscillations with Planck 2015 Data}
\author{Rabia Saleem
\thanks{rabiasaleem@ciitlahore.edu.pk}\\
Department of Mathematics,\\ COMSATS, Institute of Information
Technology Lahore, Pakistan.}

\date{}
\maketitle

\begin{abstract}
This paper is aimed to study the compelling issue of cosmic inflation during rapid
oscillations using the framework of non-minimal derivative coupling. To this end,
an anisotropic and homogeneous Bianchi I background is considered. In this context,
I developed the formalism of anisotropic oscillatory inflation and found some
constraints for the existence of inflation. In this era, the parameters related
to cosmological perturbations are evaluated, further, their graphical trajectories
are presented to check the compatibility of the model with the observational data
(Planck 2015 probe).
\end{abstract}

\section{Introduction}

In 1981, Guth (1981) invented the term ``cosmological inflation", a compelling research aspect
in modern cosmology. Inflation is an additional idea in hot big-bang (HBB) theory, which
is applied on very initial stage of the cosmic evolution. It is supposed that scale of
inflation to be long since over and the standard expansion restored, in order to maintain
the significant successes, such as the cosmic microwave background radiation (CMBR) and
nucleosynthesis. Regardless of all of its triumphs, there are some unsatisfactory issues
with HBB theory which cultivated inflation (Starobinsky 1980).

The first issue is the ``flatness problem" (an easiest one to understand): for flat universe,
$\Omega=1$ on the time scale (where $\Omega=\frac{\rho}{\rho_c};~\rho_c$ is the critical density).
In standard big-bang model, the curvature term $(aH)^{2}$ ($a,~H$ be the scale factor and Hubble
parameter) always a decreasing function of time leading to $\Omega(t)$ different from unity due
to cosmic expansion. However, according to recent observations, the value of $\Omega(t)$ is near
to unity, thus it must be same (very close to one) in the early-time. For example, its value at
\textit{Planck time} is ``$|\Omega{(t_{Pl})}|<\mathcal{O}(10^{-64})$" while ``$|\Omega{(t_{nucleo})}|
<\mathcal{O}(10^{-16})$" (during \textit{nucleosynthesis}). These values represent that there was
a need of highly fine-tuning of ``initial conditions". The inaccurate choice of ``initial conditions"
leads to the cosmos which either soon expands before the formation of structure or quickly collapses.
This dubbed as ``flatness problem" (Liddle 2000).

The ``horizon problem" illustrate that ``Why the temperature of CMBR appears the same in all directions?"
In east and west directions, the exactly same temperature of CMBR is detected, while the radiation coming from
the east and west are detached by ``$28$ billion light years". As we know that information always transformed with a
speed less than the ``speed of light", hence neither the radiation detected from two directions of the the universe
could be in thermal contact nor the regions ever have been in link. The only possibility for two regions to be in
thermal equilibrium is that they must be enough close to communicate with each other. Then, how thermal equilibrium
between two regions was attained if there was no causal connection? (Liddle 2000).

The mechanism of inflation is proposed to resolve the standard shortcomings of HBB model. ``Stripped to its bare bones",
inflation is an era of the cosmic evolution where the scale factor $a(t)$ was growing exponentially $(\ddot{a}<0)$.
The acceleration equation immediately implies that ``$\rho+3P<0$", since density is always assumed positive, so to satisfy
the inequality, pressure should be negative $(P<-\frac{\rho}{3})$. Fortunately, the symmetry breaking (concept of modern
particle physics) give ways through which this negative pressure can be achieved. A cosmos possessing $(\Lambda)$ (the
cosmological constant, representing by $P=-\rho$) is one of the typical example of inflationary cosmic expansion.
After a passage of time, the energy of $\Lambda$ decayed into ordinary matter which leads to a ``graceful exit" from
inflation and again preserve the HBB model. Unfortunately, $\Lambda$ is proved to be a very ad hoc technique. A successful
inflationary model must possesses a feasible hypothesis for the source of $\Lambda$ and a ``graceful exit" from the inflation
(Linde 1990).

The ``phase transitions" is a basic idea to obtain inflation. This is especially a dramatic event in the cosmic time-line,
a time when universe really changes its properties. It is fact that the current cosmos have passed through a series
of phase transitions as it cooled down. A curious form of matter named scalar field is consider to be responsible
for the cosmic phase transitions. It possesses negative pressure (an effective $\Lambda$) and satisfy the condition
($p+3p<0$), necessary to attain inflation. At the end of phase evolution, the inflaton (scalar particle which produced
inflation) decomposed and the inflation terminates, hopefully having attained the required cosmic expansion (by a factor of
$10^{27}$ or more).

Inflation resolves the ``flatness problem" as: ``consider a balloon being very quickly blown up
(say to the size of the sun), its surface would then look flat to us. The crucial difference
between inflation and the standard HBB model is that the size of the region of the observable
universe (given roughly by the ``Hubble length" $(cH^{-l}$; $H^{-l}$ be the age of the universe
and $c$ be the maximum speed), does not change while this happens. So, soon you are unable to
notice the curvature of the surface. While in the big bang scenario the distance you can
see increases very rapidly than the balloon expands, so you can observe more of the curvature
as time goes by." The solution of the ``horizon problem" can be described in a precise way as:
``inflation enlarges the  size of a portion of the cosmos, while keeping its peculiar scale
(the Hubble scale) fixed. This statement yields that a small patch of the universe, will be
small enough to obtain thermal contact before inflation, can expand to be much larger than the
size of our presently observable universe. Then the CMBR coming from opposite sides of the sky
really are at the same temperature because they were once in equilibrium. Equally, this provides
the opportunity to generate irregularities in the universe which can lead to structure formation"
(Linde 1990).

The oscillatory (cyclic) universe have a long past in the field of cosmology (Tolman 1934). Formerly, one of their
basic interest was that the initial conditions could in principle be evaded. However, a detailed analysis
of such models affirmed severe problems in their development in the framework of general relativity (GR).
Aside from entropy constraints that reduced the number of bounces in the early time, the basic difficulty
is the classical treatment when any bounce is singular, thereby leading to the failure of GR. Current
progress in M-theory inspired braneworld models have reborn interest in cyclic (oscillatory) universe,
despite the fact, problems still attached with these models in constructing a successful analysis of
the bounce (Khoury et al. 2001; 2002; Steinhardt and Turok 2002). An oscillating universe that subsequently
underwent an inflationary cosmic expansion after a
finite number of cycles has also been discussed (Kanekar, Sahni and Shtanov 2001). Yet, a physical method to bring about the bounces
was not implemented in this model.

Damour and Mukhanov (1998) are the pioneers of ``oscillating inflation". They proposed that, after
the slow-roll, inflation may lasts in rapid coherent oscillation during the reheating regime. Liddle
and Mazumder (1996) formulated the corresponding ``number of e-folds". The decay of scalar fields
in the oscillations to inflaton was also discussed briefly by Bartruma et al. (2014). The adiabatic
perturbation in the oscillatory inflation are investigated in (Taruya 1999). The authors in (Lee et al. 1999) extended
the work of Damour and Mukhanov (1998), by taking a coupling between the Ricci scalar curvature and inflaton.
An important form of the potential, which is needed to end the oscillatory inflation, was formulated in (Sami 2003).
The rapid oscillatory phase gives a less ``number of e-folds", so it is not possible to avoid the slow-roll phase
during this formalism. Because of few ``number of e-folds", a deep analysis of the growth of quantum fluctuations
has not been executed. To solve this difficulty, we can assume a ``non-minimal derivative coupling model".
A mess of literature exists to study the cosmological aspects of this model (Sushkov 2009; Saridakis and Sushkov 2010;
Sadjadi 2011; Yang, Gao and Gong 2015; Cai and Piao 2016; Huang and Gong 2016).

The oscillatory inflation with ``non-minimal kinetic coupling" is presented in (Sadjadi and Goodarzi 2014) which solves the issue of
few number of e-folds coming in (Damour and Mukhanov 1998) (``non-minimal derivative coupling model") as it increases the
``number of e-folds" during high-friction era. However, it is not clear from this scenario how reheating
occurs or the universe becomes radiation dominated after the end of inflation. Sadjadi and Goodarzi (2014) analyzed
the compatibility of the perturbed parameters like scalar(tensor) perturbations, power spectra and
spectral index for scalar(tensor) modes in oscillatory inflation with Planck 2013 data. The isotropic universe is
just an perfect realization to the cosmos we observe as it ignores all the structure and other observed anisotropies,
e.g., in the CMB temperature (Russell, Kilinc and Pashaev 2014). One of the great triumphs of inflation is to have a naturally embedded
mechanism to account for these anisotropies. Sharif and Saleem (2014; 2015) studied warm vector inflation in
``locally rotationally symmetric Bianchi type I" (LRS BI) universe model and verified its compatibility with WMAP7 data.

Motivated by the combined work of Sadjadi and Goodarzi (2014), I have discussed inflationary scenario during
rapid oscillation of a scalar field in non-minimal derivative coupling model. To this end, the framework of
LRS BI universe model is used. The paper is organized as follows. The basic formalism of oscillatory inflation
in the background of LRS BI universe model is given in section \textbf{2}. Section \textbf{3} deals with
cosmological perturbations during minimal and non-minimal cases. I evaluate explicit expressions for
perturbed parameters and analyzed them through graphical trajectories by constraining the model parameters with Planck 2015
observations. Finally, the results are concluded in the last section.

\section{Formalism of Anisotropic oscillatory Inflation}

In the past three decades, various inflationary models have been proposed,
where in many of them inflation is driven by a canonical scalar field $(\phi)$,
rolling slowly in an almost flat potential. Higgs boson is considered to be a
natural candidate for inflaton (Bezrukov and Shaposhnikov 2008; Bezrukov et al. 2011). Inspired by this idea, Germani and
Kehagias (2010) introduced a non-minimal coupling between kinetic term of $\phi$
and the Einstein tensor $(G^{\mu\nu})$, tried to consider the inflaton as the Higgs
boson, without violating the unitarity bound. This model is specified by the following
action
\begin{equation*}
S=\int\left(\frac{M^{2}_{P}}{2}R-\frac{1}{2}\Delta^{\mu\nu}\partial_{\mu}\phi
\partial_{\nu}\phi-V(\phi)\right)\sqrt{-g}d^4x,
\end{equation*}
where $\Delta^{\mu\nu}=g^{\mu\nu}-\frac{1}{M^2}G^{\mu\nu}$ ($M$ be a coupling
constant with dimension of mass, $G^{\mu\nu}=R^{\mu\nu}-\frac{1}{2}Rg^{\mu\nu}$)
and $M_P=2.435\times10^{18}GeV$ is the reduced Planck mass. I have considered the
gravitational enhanced friction model in the framework of LRS BI model.
The model is represented by the following line element
\begin{equation*}
ds^2=-dt^2+a^2(t)dx^2+b^2(t)(dy^2+dz^2),
\end{equation*}
with $a(t),~b(t)$ are the scale factors along $x$-axis and
$(y,z)$-axis, respectively. This metric can be transformed in the
following form using a linear relationship $a=b^{m},~m\neq1$
(Sharif and Zubair 2010)
\begin{equation*}
ds^2=-dt^2+b^{2m}(t)dx^2+b^2(t)(dy^2+dz^2).
\end{equation*}

The equation of motion for inflaton is given as
\begin{eqnarray}\nonumber
\left(1+(m+2)\frac{H^{2}_{2}}{M^2}\right)\ddot{\phi}&+&(m+2)H_{2}\left(1+(m+2)\frac{H^{2}_{2}}{M^2}
+(m+2)\frac{2\dot{H}_2}{3M^2}\right)\dot{\phi}\\\label{1}&=&-V^{\prime}(\phi),
\end{eqnarray}
where $H_2=\frac{\dot{b}}{b},~V(\phi)$ are the directional Hubble parameter
and the effective potential, respectively. Dot and prime represent the derivative with respect to time
and scalar field. The energy density $(\rho_{\phi})$ and the pressure $(P_{\phi})$
for homogeneous and anisotropic scalar field can be expressed as, respectively
\begin{eqnarray}\nonumber
\rho_{\phi}&=&\left(1+(m+2)^2\frac{H^{2}_{2}}{M^2}\right)\frac{\dot{\phi}^2}{2}+V(\phi),\\\nonumber
P_{\phi}&=&\left(1-(m+2)\frac{H^{2}_{2}}{M^2}
-(m+2)\frac{2\dot{H}_2}{3M^2}\right)\frac{\dot{\phi}^2}{2}-\frac{2(m+2)}{3M^2}\dot{\phi}\ddot{\phi}-V(\phi).\\\label{2}
\end{eqnarray}
The dynamics of anisotropic oscillatory inflation is described by the evolution equation given by
\begin{equation}\label{3}
H^{2}_{2}=\frac{1}{(1+2m)M^2_{P}}\rho_{\phi}.
\end{equation}

Here, I consider the rapid oscillatory solution for $\phi$, with both time dependent amplitude
$\Phi(t)$ (the highest point of oscillation at which $\dot{\phi}=0$) and frequency
$\omega(t)=\frac{1}{T(t)}$, where $T(t)$ (the period of oscillation) is defined as
\begin{equation}\label{4}
T=2\int^{\phi}_{-\phi}\frac{d\phi}{\dot{\phi}}.
\end{equation}
The rapid oscillation phase obeys the following conditions
\begin{equation}\label{5}
H_{2}(t)\ll\frac{1}{T(t)};\quad |\frac{\dot{H}_{2}}{H_{2}}|\ll\frac{1}{T(t)},
\end{equation}
implying that the directional Hubble parameter changes insignificantly during
one oscillation (i.e., $H_{2}(t^{\prime})\approx H_{2}(t)$ for $t\leq t^{\prime}\leq t+T(t)$).
Equations (\ref{3}) and (\ref{5}) together yield that similar to $H_{2}$,
$\rho_{\phi}$ also remains approximately constant during one period. We take the constant value of
inflaton density at the amplitude, $\Phi$ (where $\dot{\phi}\mid_{|\phi|=\Phi}=0$).
Therefore $\rho_{\phi}$ during one oscillation can be expressed in terms of
$V(\phi)$ as $\rho_{\phi}=V(\phi)$ at the corresponding $\Phi$ (Shtanov, Traschen and Brandenberger 1995).

Therefore, for a power law potential, I expect that $|\frac{\dot{\phi}}{\phi}|\ll\frac{1}{T}$.
To elucidate more this subject, the rapid oscillating scalar field is depicted numerically using
Eqs.(\ref{1})-(\ref{3}) for a quadratic potential, showing that the amplitude of oscillation changes
very slowly during one oscillation. Also in Fig. \textbf{1}, the oscillation of the scalar field for
the potential $V(\phi)=\lambda|\phi|^{q}$ is numerically shown for $q\in(-2,\infty)$ (the reason for this choice
will be revealed when we will determine our parameters from astrophysical data in the third section).

The adiabatic index $(\gamma)$ is related with equation of state (EoS) parameter
$(w=\frac{P_\phi}{\rho_\phi})$ as $w=w+1$. In the rapid oscillation phase, it can be
calculated using Eqs.(\ref{3}), (\ref{4}) and the expression $\rho_{\phi}=V(\phi)$ as follows
\begin{eqnarray}\nonumber
\gamma&=&\frac{\langle P_\phi+\rho_\phi\rangle}{\langle\rho_\phi\rangle}=\frac{1}{\langle\rho_\phi\rangle}
\left[\left(1+\frac{(m+2)^2H^{2}_{2}}{3M^2}\right)\langle\dot{\phi}^2\rangle
-\frac{d}{dt}\left(\frac{(m+2)H_{2}\dot{\phi}^2}{3M^2}\right)\right],\\\nonumber&=&
\left(1+\frac{(m+2)^2H^{2}_{2}}{3M^2}\right)\frac{\langle\dot{\phi}^2\rangle}{\langle\rho_\phi\rangle}
=\frac{2\left(1+\frac{(m+2)^2H^{2}_{2}}{3M^2}\right)}{\left(1+\frac{(m+2)^2H^{2}_{2}}{M^2}\right)}
\frac{\langle\rho_{\phi}-V(\phi)\rangle}{\langle\rho_\phi\rangle},\\\label{6}&=&
\frac{2\left(1+\frac{(m+2)^2H^{2}_{2}}{3M^2}\right)}{\left(1+\frac{(m+2)^2H^{2}_{2}}{M^2}\right)V(\Phi)}
\frac{\int^{\phi}_{-\phi}\sqrt{V(\Phi)-V(\phi)}}{\int^{\phi}_{-\phi}\frac{d\phi}{\sqrt{V(\Phi)-V(\phi)}d\phi}},
\end{eqnarray}
where the bracket $\langle\cdots\rangle=\frac{\int^{t+T}_{t}\cdots dt^{\prime}}{T}$ denotes the time
averaging. To obtain above equation, it is taken into account the fact that $\dot{\phi}$ vanishes at
$\mid\phi\mid=\Phi$. Using power law potentials of the form $V(\phi)=\lambda\phi^q ~(\lambda\in\mathcal{R})$,
one can easily obtained the average adiabatic index as
\begin{equation}\label{7}
\gamma=\frac{2q}{q+2}\left(\frac{1+\frac{(m+2)^2H^{2}_{2}}{3M^2}}{1+\frac{(m+2)^2H^{2}_{2}}{M^2}}\right),
\end{equation}
where $q$ is a dimensionless real parameter, while the limit $q\rightarrow0$, gives a logarithmic potential.
In the minimal coupling case $(M\rightarrow\infty)$, the adiabatic index reduced to $\gamma=\frac{2q}{q+2}$.
In the high friction regime, where $\frac{H^{2}_{2}}{M^2}\gg1\Rightarrow\gamma=\frac{2q}{3q+6}$.

The average rate of change of $\rho_{\phi}$ is given as
\begin{equation}\label{8}
\langle\dot{\rho}_{\phi}\rangle=\lim_{\quad\quad\quad T\rightarrow0}\frac{\rho_{\phi}(t+T)-\rho_{\phi}(t)}{T}\simeq\dot{\rho}_{\phi}.
\end{equation}
By taking the average of the continuity equation, we obtain
\begin{equation}\label{9}
\langle\dot{\rho}_{\phi}+(m+2)H_2(\rho_{\phi}+P_{\phi})\rangle=\dot{\rho}_{\phi}+(m+2)H_2\gamma\rho_{\phi}=0,
\end{equation}
where $\gamma$ is given in Eq.(\ref{6}). For constant $\gamma$, the Eqs.(\ref{3}) and (\ref{9})
can be solved analytically. For the situation $\frac{H^{2}_{2}}{M^2}\gg1$, the analytical solutions
for $\rho _{\phi},~b(t)$ and $H_2(t)$ are as follows, respectively
\begin{equation}\label{10}
\rho _{\phi}\propto b^{-(m+2)\gamma},\quad b\propto t^{2/(m+2)\gamma},
\quad H_{2}=\frac{2}{(m+2)\gamma t}.
\end{equation}

Next, I restrict the work to the high friction regime to find out the analytical solution of the model.
In the case of power law potential, the period is calculated by the following formula
\begin{eqnarray}\nonumber
T&=&2\int^{\Phi}_{-\Phi}\frac{d\phi}{\dot{\phi}},\\\nonumber&=&2\sqrt{\frac{(m+2)H^{2}_{2}}
{2M^2}}\int^{\Phi}_{-\Phi}\frac{1}{\sqrt{\rho_{\phi}-V(\phi)}}d\phi,\\\nonumber&=&2\sqrt{\frac{(m+2)H^{2}_{2}}
{2M^2}}\int^{\Phi}_{-\Phi}\frac{1}{\sqrt{\lambda\Phi^{q}-\lambda\phi^q}}d\phi,
\\\nonumber&=&2\sqrt{\frac{\pi(m+2)}{2M^2\lambda}}
\frac{\Gamma(\frac{1}{q})}{q\Gamma(\frac{q+2}{2q})}H_2\Phi^{\frac{2-q}{2}},
\\\label{11}&=&\frac{2}{MM_{p}}\sqrt{\frac{\pi(m+2)}{2(1+2m)}}
\frac{\Gamma(\frac{1}{q})}{q\Gamma(\frac{q+2}{2q})}\Phi,
\end{eqnarray}
where $H_{2}=\sqrt{\frac{\lambda}{(1+2m)}}\frac{\Phi^{\frac{q}{2}}}{M_{p}}$.
\begin{figure}
\center\epsfig{file=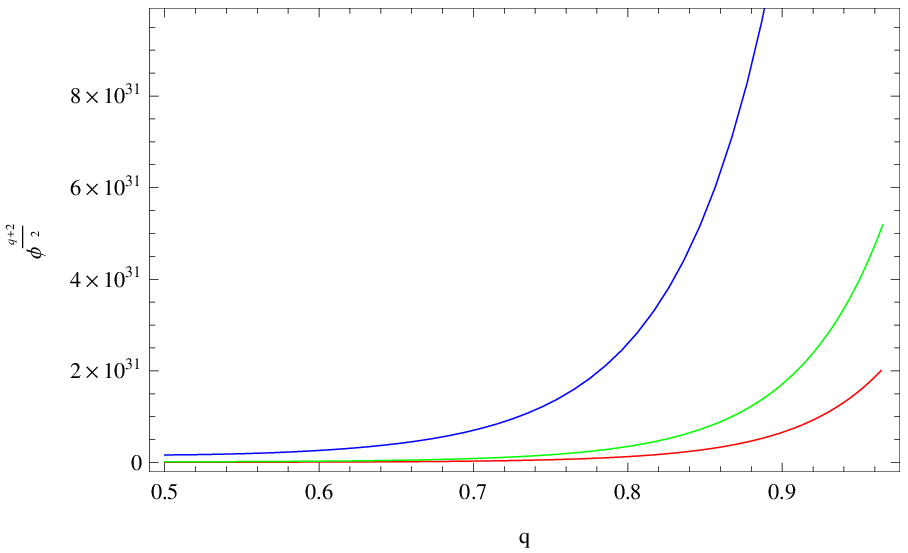,
width=0.50\linewidth}\epsfig{file=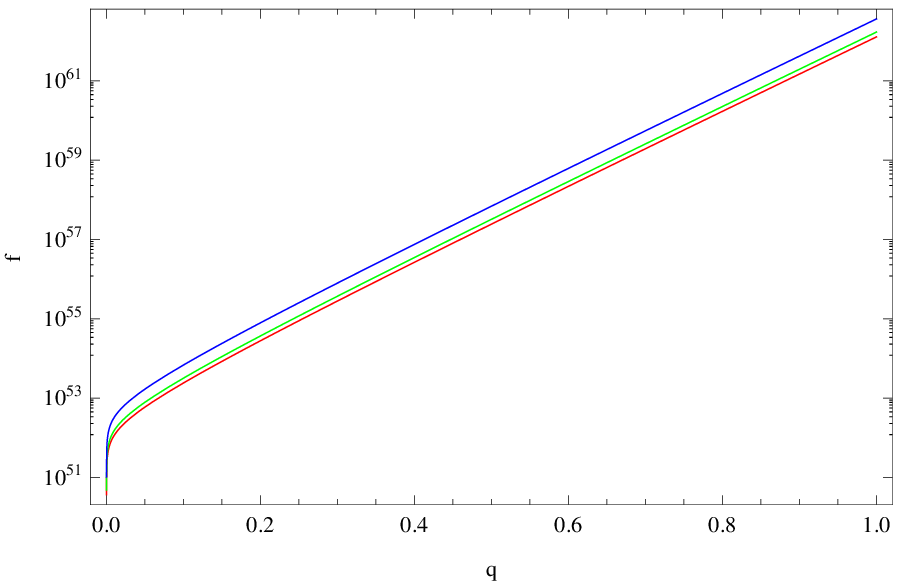,
width=0.50\linewidth}\caption{(left) The behavior of $\phi^{\frac{q+2}{2}}$ versus $q$; (right) $f$
versus $q$ are plotted for $m=1.5$ (red); $m=2.5$ (green); $m=10$ (blue).}
\end{figure}
The condition, $H_{2}T\ll1$, can be rewritten in terms of inflaton using above two
expressions as
\begin{equation}\label{12}
\Phi^{\frac{q+2}{2}}\ll\frac{MM_{p}^{2}(1+2m)q}{\sqrt{2\pi\lambda(m+2)}}
\frac{\Gamma(\frac{q+2}{2q})}{\Gamma(\frac{1}{q})},
\end{equation}
where the scale $M$ present in the above expression reduces the scale of $\phi$ as compared to
the minimal case, which gives
\begin{equation}\label{13}
\Phi\ll\frac{\sqrt{(1+2m)}q}{\sqrt{2\pi(m+2)}}
\frac{\Gamma(\frac{q+2}{2q})}{\Gamma(\frac{1}{q})}M_{p}.
\end{equation}
Therefore, the evaluated solution must be valid in the domain of Eq.(\ref{12})
which specifies a bound for $H_2$ and consequently for $\rho_\phi$
during rapid oscillation. In order to check, either the inequality given in Eq.(\ref{12})
holds or not, we have plotted Fig.\textbf{1}. The left term, $\Phi^{\frac{q+2}{2}}$ and expression on
right hand side (say $f$) of Eq.(\ref{12}) are plotted versus $q$, for specified values of $m$,
in the left and right panel of Fig.\textbf{1}, respectively. It is very much clear from the comparison
of left and right graph of Fig.\textbf{1} that the value of $\Phi^{\frac{q+2}{2}}$ is much less
than the expression $f$ for $q>0$ and $m>0;~m\neq1$ (the result also holds for $q<0$ but for stable model,
I am restricting my work to $q>0$).

The slow-roll conditions, $\ddot{\phi}\ll(m+2)H_2\dot{\phi}$ and $\rho_{\phi}\approx V(\phi)$
together with the expression of $\rho_{\phi}$ given in the first equation of Eq.(\ref{2}) generates
the following inequality
\begin{equation}\label{14}
\left(1+\frac{(m+2)^2H^{2}_2}{M^2}\right)\frac{\dot{\phi}^2}{2}\ll V(\phi).
\end{equation}
In high friction regime ($\frac{H^{2}_2}{M^2}\gg1$), all the above mentioned conditions are satisfied
when $\phi^{q+2}\gg\frac{q^2M^2M^{4}_p}{\lambda}$ (opposite to Eq.(\ref{12})) holds. Equation (\ref{14})
leads to
\begin{equation}\label{15}
\frac{(m+2)^2H^{2}_2}{2M^2}{\dot{\phi}}^2\ll V(\phi)\sim(1+2m)M^{2}_pH^{2}_2,
\end{equation}
resulting that
\begin{equation}\label{16}
{\dot{\phi}}^2\ll\frac{2(1+2m)}{(m+2)^2}M^2M^{2}_p.
\end{equation}
While during quasi periodic regime, Eqs.(\ref{3}) and (\ref{6}) produce the expression
$\langle{\dot{\phi}}^2\rangle\approx\gamma M^2M^{2}_p$.

Inflation takes place when $\ddot{b}>0$ or equivalently $\gamma<\frac{2}{3}$, putting in the expression
of $\gamma$, I can fix a range of $q\in(-2,\infty)$. While during the minimal case ($\gamma=\frac{2q}{q+2}$), inflation occurs
only for the short interval $q\in(-2,1)$. As already mentioned that the inflation continues as long as $\gamma<\frac{2}{3}$,
then fourth equality of Eq.(\ref{6}) produces
\begin{equation}\label{17}
\left(\frac{1+\frac{(m+2)^2H^{2}_2}{3M^2}}{1+\frac{(m+2)^2H^{2}_2}{M^2}}\right)\langle\rho_\phi-V(\phi)\rangle
<\frac{1}{3}\langle\rho_\phi\rangle.
\end{equation}
The above expression leads to constraint the potential in high friction and minimal regimes as, respectively
\begin{equation}\nonumber
\frac{H^{2}_2}{M^2}\gg1\Rightarrow\langle V(\phi)\rangle\gg0, \quad\frac{H^{2}_2}{M^2}\rightarrow0\Rightarrow
\langle V(\phi)\rangle>\frac{2}{3}\langle \rho_(\phi)\rangle.
\end{equation}
Inflation ends also for more complicated potential such as the potential suggested by Damour-Mukhanov (1998)
\begin{equation}\label{15}
V(\phi)=\nu\left(\left(\frac{\phi^2}{\phi^{2}_c}+1\right)^{\frac{q}{2}}-d\right),
\end{equation}
where $d>0$ is a dimensionless real number while $\nu,~\phi_c$ are real parameters with dimensions $[mass]^4$
and $[mass]$, respectively. For $\phi\gg\phi_c$, the above potential reduced to simple power law potential.

The average potential can be evaluated by the following formula (Sami 2003)
\begin{equation*}
\langle V(\phi)\rangle=\frac{\int^{\Phi}_{-\Phi}\frac{V(\phi)}{\dot{\phi}}d\phi}{\int^{\Phi}_{-\Phi}
\frac{d\phi}{\dot{\phi}}},
\end{equation*}
leads to find that the inflation continues as long as $\int^{\Phi}_{-\Phi}V(\phi)d\phi>0$. For the potential,
given in Eq.(\ref{15}), we have
\begin{equation}\label{17}
\int^{1}_{-1}((\mathfrak{b}^2\chi^2+1)^{\frac{q}{2}}-d)d\chi>0,
\end{equation}
here, $\mathfrak{b}=\frac{\Phi}{\phi_c}$ and $\chi=\frac{\phi}{\Phi}$. The above equation results in
that the inflation continues whenever $d<g(\mathfrak{b},q)$, where
\begin{eqnarray}\label{18}
g(\mathfrak{b},q)&=&\frac{G(\mathfrak{b},q)}{2\mathfrak{b}\Gamma(\frac{q+3}{2})(1+q)\Gamma(-\frac{q}{2})},\\\nonumber
G(\mathfrak{b},q)&=&-\pi^{\frac{3}{2}}(q+1)sec(\frac{\pi q}{2})
_2F_1(-\frac{q}{2};-\frac{1+q}{2};\frac{1-q}{2};-\mathfrak{b}^{-2})
\Gamma(-\frac{q}{2})\Gamma(-\frac{q+3}{2}),
\end{eqnarray}
where $_2F_1$ is the Gauss hypergeometric function. When this inequality is violated (such that
$d=g(\mathfrak{b}_{end},q)$ and $d>g(\mathfrak{b}(t>t_{end}),q)$), inflation ceases at $t_{end}$. The graphical analysis
of the function is shown in Fig. \textbf{2} for three different values of $\mathfrak{b}=1,~5,~10$. It is observed
that $g(\mathfrak{b},q)$ has an increasing behavior for all values of $\mathfrak{b}>0$ in the range $q\in(-\infty,1)$
(in agreement with the recent values of the parameters evaluate in the next section). The point of intersection
of the three curves is $(1,0)$, so inflation ends for $d>1$ and $d\simeq1~ (\phi\sim\phi_c)$.
\begin{figure}
\center\epsfig{file=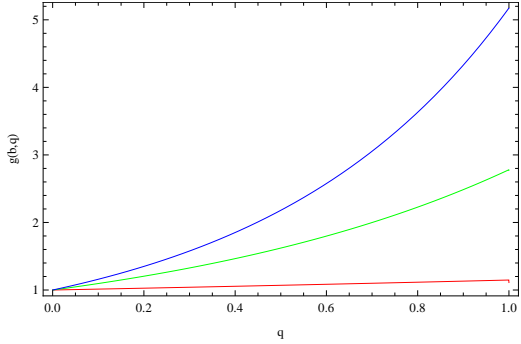,
width=0.50\linewidth}\caption{$g(\mathfrak{b},q)$ versus $q$:
for $\mathfrak{b}=1$ (red); $\mathfrak{b}=5$ (green); $\mathfrak{b}=10$ (blue).}
\end{figure}

The number of e-folds $(\mathcal{N})$ from a specific time $(t_{\ast})$ until the end of inflation
$(t_{end})$ can be defined as (Liddle and Mazumder 1996)
\begin{equation}\label{19}
\mathcal{N}=\ln\frac{b_{end}{H_2}_{end}}{b_{\ast}{H_2}_{\ast}},
\end{equation}
here, $\mathcal{N}$ is a measure of $\ln(bH_2)$, increases during inflation. Substituting
the expressions of scale factor and directional Hubble parameter from Eq.(\ref{10}), we arrive at
\begin{equation}\label{20}
\mathcal{N}=\frac{q}{2}\left(\frac{2}{(m+2)\gamma}-1\right)\ln\frac{\phi_{\ast}}{{\phi}_{end}}.
\end{equation}
During high friction and minimal regimes, the above expression turns out to be as
\begin{eqnarray}\nonumber
\mathcal{N}&=&\frac{3}{2}\left(\frac{q-m}{m+2}\right)\ln\left(\frac{\phi_{\ast}}{{\phi}_{end}}\right),\\\label{21}
\mathcal{N}_{min}&=&\left(\frac{q-2m-2}{2(m+2)}\right)\ln\left(\frac{\phi_{\ast}}{{\phi}_{end}}\right),
\end{eqnarray}
respectively. On comparing, we can note that our considered model has more number of e-folds as compared to
$\mathcal{N}_{min}$ with a common term $\ln\left(\frac{\phi_{\ast}}{{\phi}_{end}}\right)$. During intermediate
regime (where high friction condition violates), we can not obtain feasible solution for $b$ and $H_2$,
hence unable to conclude a simple form for $\mathcal{N}$.

\begin{figure}
\center\epsfig{file=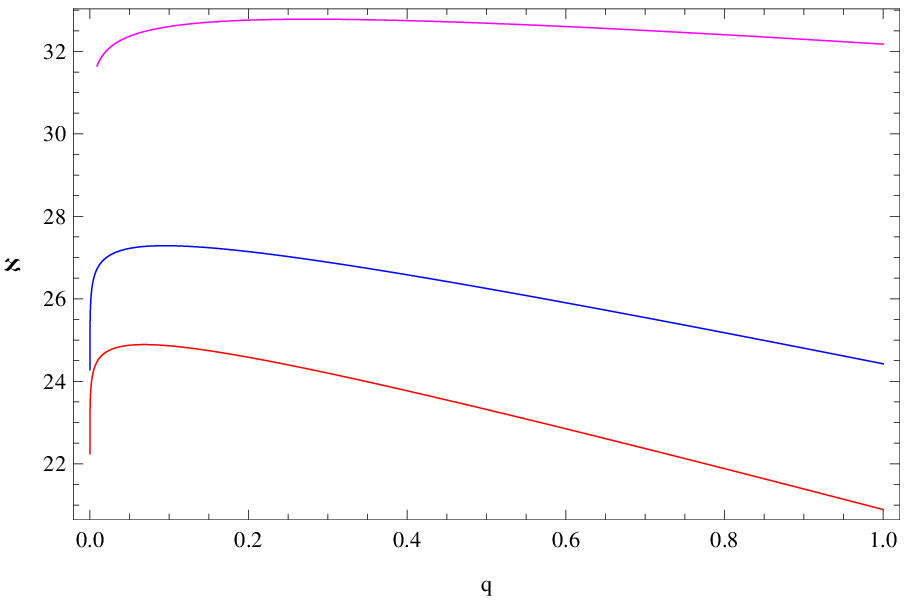,
width=0.50\linewidth}\epsfig{file=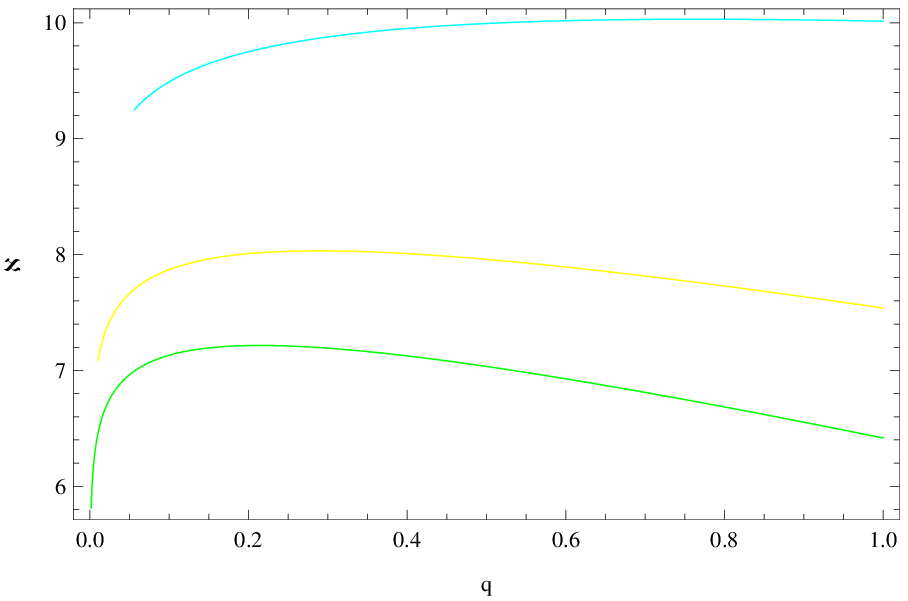,
width=0.50\linewidth}\caption{(left) $\mathcal{N}$ versus $q$ for $\Phi_{end}\sim10^{-17}m_P,~m=1.1$
(red); $m=2.5$ (blue); $m=10$ (purple); (right)
$\mathcal{N}$ versus $q$ for $\Phi_{end}\sim10^{-6}m_P,~m=1.1$ (green); $m=2.5$ (yellow); $m=10$ (cyan).}
\end{figure}
Now, we specify a lower bound for $\mathcal{N}$ during inflation. Let us take $t_k$ be the time where
$\lambda_{k}=\frac{1}{k}$ (length scale), attributed to the wavelength $k=\frac{1}{\lambda_{k}}
=b(t_k)H_2(t_k)$ (where $b(t_0)=1$ and $t_0$ be the present time), exited the Hubble radius during inflation.
The LSS observations are limited to scales of about $1Mpc$ (denoted by $\lambda_{min}$) to the present
Hubble radius ($\lambda_{max}$). These observable scales crossed the Hubble radius during the following visible
e-folding
\begin{equation}\label{22}
\mathcal{N}_{vis}=\ln\left(\frac{\lambda_{max}}{\lambda_{min}}\right)=\ln\left(\frac{H^{-1}_0}{1Mpc}\right).
\end{equation}
Putting $H_0=67.3km/sMpc^{-1}$ $(95\% C.L.)$ (Ade et al.2014a; 2014b), we obtain $\mathcal{N}_{vis}=8.4$. Hence, all relevant scales exited the Hubble
radius during $8.4$ e-folding after $\frac{1}{H_0}$'s exit, so $\mathcal{N}>8.4$.
Equations (\ref{13}) and (\ref{21}) lead to following e-folding number during minimal case
\begin{equation}\nonumber
\mathcal{N}_{min}<\left(\frac{q-2m-2}{2(m+2)}\right)\ln\left(\left(\frac{\sqrt{2(1+2m)}}{2\sqrt{\pi(m+2)}}
\frac{q\Gamma(\frac{q+2}{2q})}{\Gamma(\frac{1}{q})}\right)\frac{M_P}{\phi_{end}}\right),
\end{equation}
where $\phi_{end}$ depends on the chosen potential (given in Eq.(\ref{15})). For instance, $\Phi_{end}$
is of the same order of $\phi_{c}$ for $d=1$ in Eq.(\ref{15}). To set an upper bound on $\mathcal{N}_{min}$,
I have plotted trajectories for $\mathcal{N}-q$ taking $\Phi_{end}$ is equivalent to electroweak scale, i.e.,
$\Phi_{end}\sim10^{-17}m_P$ ($m_P$ be the planck mass) (left panel) and $\Phi_{end}\sim10^{-6}m_P$ (right panel)
and varying $m=1.5,~2.5,~10$ in Fig.\textbf{3}. It is noticed from both graphs of Fig.\textbf{3} that an increment
in the scale of $\Phi_{end}$ leads to decrease the value of $\mathcal{N}_{min}$. An increasing relationship also observed
between $\mathcal{N}$ and $m$. The electroweak scale sets an upper bound, i.e., $\mathcal{N}_{min}<24~(m=1.5),~\mathcal{N}_{min}<26~(m=2.5),~\mathcal{N}_{min}<32~(m=10)$ while $\mathcal{N}_{min}<7~(m=1.5),~\mathcal{N}_{min}<8~(m=2.5),~\mathcal{N}_{min}<10~(m=10)$ for $\Phi_{end}\sim10^{-6}m_P$.
The theory may become more viable at least in the context of perturbations generation as the anisotropic model
has ability to provide more number of e-folds ($\mathcal{N}>8.4$) in non-minimal case
as compared to the minimal case.

Again considering the scale $k=\frac{1}{\lambda_{k}}=b(t_k)H_2(t_k)$, a wave-number had chance to exit the
Hubble radius during rapid oscillatory phase, following the condition $k\ll\frac{1}{T(t_k)}$
($H_2T\ll 1;~b(t_0)=1$). To examine this condition, we proceed as that the maximum scale of our
observable universe is of the same order of magnitude as $\lambda_{max}=\frac{1}{H_2}$.
Since an upper bound for period $T$ can be determined using last equality of Eq.(\ref{11}) and
Eq.(\ref{13}) during rapid oscillation: $T(t)<T_u$. Therefore, $H_2T_u\ll1$ guarantees the
consistency of our assumptions with the horizon exit of $\lambda_{max}$ during rapid oscillation era.
This can be elaborated as
\begin{equation}\label{23}
H_{2}<\left(\frac{q-2m-2}{2(m+2)}\right)\ln\left(\frac{\sqrt{2(1+2m)}}{2\sqrt{\pi(m+2)}}
\frac{q\Gamma(\frac{q+2}{2q})}{\Gamma(\frac{1}{q})}\right)\frac{M_P}{\phi_{end}}.
\end{equation}
If the anisotropic model provides enough e-folds $(N>N_{vis})$ after this exit, also holds the
above constraint, then, we can claim that other large cosmological observable scales had also the
possibility to exit the Hubble radius during this stage of inflation. Next, I will study
perturbations generation and check model parameter's compatibility based on recent astrophysical data.

\section{Anisotropic Cosmological Perturbations}

In order to study the scalar and the tensor fluctuations, I decouple the spacetime into two components:
the background and the perturbations. Further, I have considered homogeneous and anisotropic LRS BI background
corresponding to the rapid oscillatory inflation in the context of non minimal derivative coupling model
studied in the last section. The Mukhanov-Sasaki equation is used in order to analyze the quantum
perturbations in rapid oscillation era. This equation, for scalar and tensor perturbations in
non-minimal derivative coupling model, is written as follows (Germani and Watanabe 2011)
\begin{equation}\label{24}
\frac{d^2v_{(s,t)}k}{d{\eta}^2}+\left(c^2_{(s,t)}k^2-\frac{1}{z_{(s,t)}}\frac{d^2
z_{(s,t)}}{d\eta^2}\right)v_{(s,t)}k=0,
\end{equation}
where $c_{s},~c_{t}$ are the speed of sound associated with scalar and tensor modes, respectively.
The other terms involved in the above equation like the conformal time $(\eta)$ and $z_s,~z_t$
are defined as follows
\begin{eqnarray}\nonumber
\eta(t)&=&\int^{t}\frac{dt^{\prime}}{b^{\frac{m+2}{3}}(t^{\prime})},\quad
z_s=b^{\frac{m+2}{3}}(t)\left(\frac{3}{m+2}\right)\frac{M_p\xi}{H_2}\sqrt{2\Sigma},\\\nonumber
z_t&=&b^{\frac{m+2}{3}}(t)M_p\sqrt{1-\alpha}\frac{\sqrt{e^{\lambda}_{ij}e^{\lambda}_{ij}}}{2},\\\label{25}
\end{eqnarray}
where
\begin{equation}\nonumber
\xi=\frac{1-\alpha}{1-3\alpha},\quad \Sigma=M^2\alpha\left[1+\frac{(m+2)^2H^2_2}{3M^2}
\left(\frac{1+3\alpha}{1-\alpha}\right)\right].
\end{equation}
Further,
\begin{equation}\nonumber
\alpha=\frac{\dot{\phi}^2}{2M^2M^2_p},\quad c^{2}_s=\frac{(m+2)^2H^2_2}{9\xi^2\Sigma}
\epsilon_s,\quad c^{2}_t=\frac{1+\alpha}{1-\alpha},
\end{equation}
where $\epsilon_{s}$ is given by
\begin{eqnarray}\nonumber
\epsilon_s&=&\frac{1}{b^{\frac{m+2}{3}}(t)}\frac{d}{dt}\left[\frac{b^{\frac{m+2}{3}}(t)\xi}
{\left(\frac{m+2}{3}\right)H_2}(1-\alpha)\right]-(1+\alpha),\\\label{26}
&=&\left(\frac{3}{m+2}\right)\frac{(1-\alpha)^2}{(1-3\alpha)}\left(1-\frac{\dot{H}_2}
{H^2_2}\right)-(1+\alpha).
\end{eqnarray}

Germani et al. (2012) studied Mukhanov-Sasaki equation for quasi-de Sitter background
during slow-roll regime, where $\alpha=0$. Since during rapid oscillation stage, $b(t)$ is a power law function
of time, therefore $\epsilon_s=-\left(\frac{3}{m+2}\right)\frac{\dot{H}_2}{H^2_2}\approx\frac{q}{q+2}$.
The second equality (constant $\epsilon_s$) is obtained using previous relationship of $H_2$ in terms of
$\gamma$ (given in Eq.\ref{10}). The expressions for $c_s$ and $c_t$ in terms of $q$ can be calculated
using $\xi,~\Sigma$ as follows
\begin{eqnarray}\nonumber
c^2_s&=&\frac{(1-3\alpha)^2}{3\alpha(1-\alpha)(1+3\alpha)}\epsilon_s=\frac{\left(1-\left(\frac{1+2m}{m+2}\right)
\left(\frac{q}{q+2}\right)\right)^2}{(\frac{1+2m}{m+2})
\left(1-(\frac{1+2m}{m+2})(\frac{q}{3q+6})\right)\left(1+(\frac{1+2m}{m+2})(\frac{q}{q+2})\right)},\\\label{27}
c^2_t&=&\frac{q(5m+7)+6(m+2)}{q(m+5)+6(m+2)}.
\end{eqnarray}
From first equality of the above equation, it is noticed that $c_s$ is also constant like $\epsilon_s$.
Figure \textbf{4} shows that the square speed of sound lies in the range $0<c_s<1$ for all values of $q>0$ 
and $m>0;~m\neq1$. The case $q<0$ leads us to generate instabilities producing negative $c^{2}_s$. 
In the rapid oscillation era, using $\xi,~\Sigma,~H_2$ in $z_s$, we get
\begin{equation}\label{28}
z_s=b^{\frac{m+2}{3}}(t)M_p\left(\frac{1-\alpha}{1-3\alpha}\right)\left(\frac{3}{m+2}\right)
\sqrt{\frac{2\alpha(m+2)^2(1+3\alpha)}{3\left(1-\alpha\right)}}.
\end{equation}
\begin{figure}
\center\epsfig{file=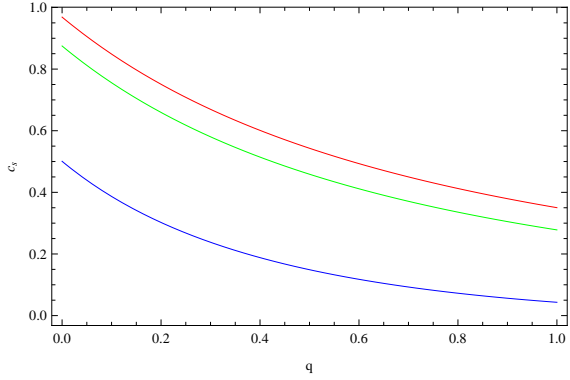,
width=0.55\linewidth}\caption{Variations in $c_s$ versus $q$:
for $m=1.1$ (red); $m=2.5$ (green); $m=10^{3}$ (blue).}
\end{figure}
Since scale factor can be written in terms of conformal time as $b(\eta)\propto
\eta^{-\left(\frac{3q+6}{2(m+2)}\right)}$, therefore, $z_{(s,t)}=\beta_{(s,t)}b(\eta)$ where
\begin{eqnarray}\nonumber
\beta_s&=&M_p\frac{\left(1-(\frac{1+2m}{m+2})(\frac{q}{3q+6})\right)}
{\left(1+(\frac{1+2m}{m+2})(\frac{q}{q+2})\right)}
\sqrt{\frac{2(m+2)(1+2m)(\frac{q}{3q+6})(1+(\frac{1+2m}{m+2})(\frac{q}{q+2}))}
{3\left(1-(\frac{1+2m}{m+2})(\frac{q}{q+2})\right)}},
\\\nonumber\beta_t&=&M_p\sqrt{1-\left(\frac{1+2m}{m+2}\right)\left(\frac{q}{3q+6}\right)}
\frac{\sqrt{e^{\lambda}_{ij}e^{\lambda}_{ij}}}{2}.
\end{eqnarray}
So, the conformal time derivative of $z_{(s,t)}$ is given by
\begin{equation}\nonumber
\frac{1}{z_{(s,t)}}\frac{d^2z_{(s,t)}}{d\eta^2}=\left(\frac{3q+6}{2(m+2)}
\right)\left(\left(\frac{3q+6}{2(m+2)}\right)-1\right)\eta^{-2}.
\end{equation}
Hence, the mode function satisfies the following differential equation
\begin{eqnarray}\nonumber
\frac{d^2v_{(s,t)}}{d\eta^2}+\left(c^2_{(s,t)}k^2-\frac{1}{z_{(s,t)}}
\frac{d^2z_{(s,t)}}{d\eta^2}\right)v_{(s,t)}k&=&0,\\\nonumber
\frac{d^2v_{(s,t)}}{d\eta^2}+\left(c^2_{(s,t)}k^2-\left(\frac{3q+6}{2(m+2)}
\right)\left(\left(\frac{3q+6}{2(m+2)}\right)-1\right)\eta^{-2}
\right)v_{(s,t)}k&=&0,
\end{eqnarray}
whose solution is
\begin{equation}\label{29}
v_{(s,t)k}(\eta)=|\eta|^{\frac{1}{2}}\left[c^{(1)}_{(s,t)}(k)\mathcal{H}^{(1)}_v(c_{(s,t)}k|\eta|)
+c^{(2)}_{(s,t)}(k)\mathcal{H}^{(2)}_v(c_{(s,t)}k|\eta|)\right],
\end{equation}
where $c^{(1)}_{(s,t)}(k)$ and $c^{(2)}_{(s,t)}(k)$ are the integrating constants while
$\mathcal{H}^{(1)},~\mathcal{H}^{(2)}$ are the Hankle functions of the first and second kind of order
$v=\frac{m+2}{2}+\frac{q}{2}$. I have used the Bunch-Davies vacuum by imposing the condition that the mode
function approaches the vacuum of the Minkowski spacetime in the short wavelength limit $\frac{b}{k}\ll\frac{1}{H_2}$,
where the mode is well with in the horizon. In the rapid oscillation epoch, we have $bH_2\propto\frac{1}{|\eta|}$ resulting $k\eta\gg1$.
Under this limit, the Bunch-Davies mode function is given by
$v_k(\eta)\approx\frac{1}{\sqrt{2c_sk}}e^{-ic_sk\eta}$.
\begin{equation}\nonumber
v_{(s,t)k}(\eta)=\frac{\sqrt{\pi}}{2}e^{i(v+\frac{1}{2})\frac{\pi}{2}}
(-\eta)^{\frac{1}{2}}\mathcal{H}^{(1)}_v(-c_{s,t}k\eta),
\end{equation}
In the limit, $\frac{k}{bH_2}\rightarrow0$, the asymptotic form of mode
function Eq.(\ref{29}) is given by
\begin{equation}\label{30}
v_{(s,t)k}(\eta)\rightarrow e^{i(\nu+\frac{1}{2})\frac{\pi}{2}}
2^{(v-\frac{m+2}{2})}\frac{\Gamma(v)}{\Gamma(\frac{m+2}{2})}
\frac{1}{\sqrt{2c_{(s,t)}k}}(-c_{(s,t)}k\eta)^{-v+\frac{1}{2}}.
\end{equation}

To obtain scalar power spectrum $P^{\frac{1}{2}}_{(s,t)}(k)$, we use
the previous calculated expressions given as
\begin{eqnarray}\nonumber
P_{(s,t)}^{\frac{1}{2}}(k)&=&\sqrt{\frac{k^3}{2\pi^2}}\mid\frac{v_{(s,t)}k}{z_{(s,t)}}\mid,\\\nonumber
&=&\sqrt{\frac{k^3}{2\pi^2}}\frac{e^{i(v+\frac{1}{2})\frac{\pi}{2}}}{B_{(s,t)}b(\eta)}
2^{(v-\frac{m+2}{2})}\frac{\Gamma(v)}{\Gamma(\frac{m+2}{2})}
\frac{1}{\sqrt{2c_{(s,t)}k}}(-c_{(s,t)}k\eta)^{-v+\frac{1}{2}}.
\end{eqnarray}
Rewriting the formula of conformal time in the following form
\begin{equation}\label{31}
\eta=\int\frac{dt}{b^{\frac{m+2}{3}}(t)}=-\frac{1}{bH_2}+\int\frac{\epsilon db}{b^2H_2}=
-\frac{1}{bH_2}\frac{1}{1-\epsilon}.
\end{equation}
The last equality in the above equation is obtained by taking the fact that $\epsilon$
is constant in the rapid oscillation era. Putting the above value of $\eta$,
the $P_{(s,t)}$ takes the following form
\begin{equation}\nonumber
P_{(s,t)}^{\frac{1}{2}}(k)=\frac{k2^{(v-1-\frac{m+2}{2})}}{\sqrt{c_{(s,t)}}\pi B_{(s,t)}b}
\frac{\Gamma(v)}{\Gamma(\frac{m+2}{2})}\left(\frac{c_{(s,t)}k}{bH_2(1-\epsilon)}\right)^{-v+\frac{1}{2}}.
\end{equation}
Moreover, at the horizon crossing scale $c_{s}k=bH_2$, the power spectrum turned out to be
\begin{equation}\nonumber
P_{(s,t)}^{\frac{1}{2}}(k)=\frac{2^{(v-1-\frac{m+2}{2})}}{\pi B_{s,t}}
\frac{\Gamma(v)}{\Gamma(\frac{m+2}{2})}\frac{H_2}{c^{\frac{3}{2}}_{(s,t)}}
(1-\epsilon)^{v-\frac{1}{2}}.
\end{equation}
\begin{figure}
\center\epsfig{file=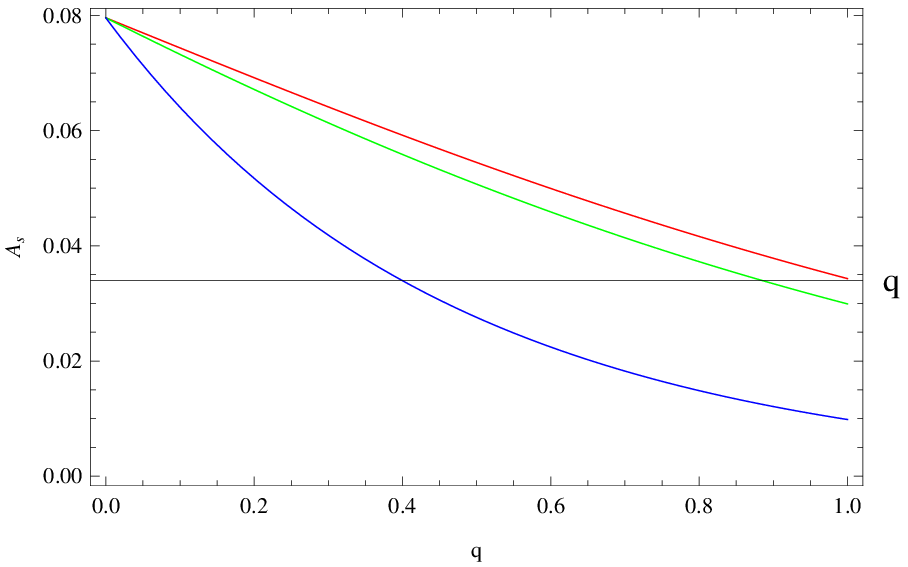,
width=0.55\linewidth}\epsfig{file=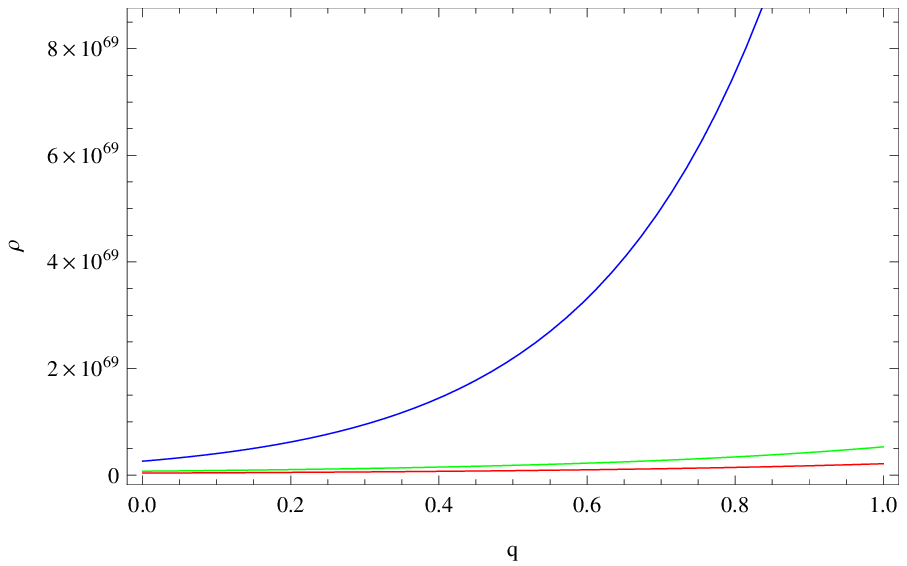,
width=0.55\linewidth}\caption{(left) The behavior of $A_s$ versus $q$; (right) $\rho$
versus $q$ are plotted for $m=1.1$ (red); $m=2.5$ (green); $m=10$ (blue).}
\end{figure}
To calculate amplitudes related to scalar and tensor spectrum, we write the above expression as
\begin{equation}\label{32}
P_{(s,t)}^{\frac{1}{2}}(k)=A_{(s,t)}(q)\frac{H_2}{M_p}\mid_{c_{(s,t)}k=bH_2},
\end{equation}
where
\begin{eqnarray}\nonumber
A_{s}(q)&=&\frac{2^{(v-1-\frac{m+2}{2})}}{\pi B_{s}c_{s}}
\frac{\Gamma(v)}{\Gamma(\frac{m+2}{2})}(1-\epsilon_s)^{v-\frac{1}{2}},\\\nonumber
A_{s}(q)&=&\frac{2^{(q-\frac{5}{2}+\frac{m+2}{2})}}{\pi(q+2)^{\frac{q}{2}-\frac{1}{2}+\frac{m+2}{2}}}
\sqrt{\frac{1-(\frac{1+2m}{m+2})(\frac{q}{q+2})}{1-(\frac{1+2m}{m+2})(\frac{q}{3q+6})}}
\frac{\Gamma(\frac{m+2}{2}+\frac{q}{2})}{\Gamma(\frac{m+2}{2})},
\end{eqnarray}
\begin{eqnarray}\nonumber
A_{t}(q)&=&\frac{2^{(q-\frac{5}{2}+\frac{m+2}{2})}}{\pi(q+2)^{\frac{q}{2}-\frac{1}{2}+\frac{m+2}{2}}}
\sqrt{\frac{q(m+5)+6(m+2)}{(1-(\frac{1+2m}{m+2})(\frac{q}{3q+6}))(q(5m+7)+6(m+2))}}\\\label{32a}&\times&
\frac{\Gamma(\frac{m+2}{2}+\frac{q}{2})}{\Gamma(\frac{m+2}{2})}.
\end{eqnarray}
The above parameters $A_{s}(q),~A_{t}(q)$ are obtained using the values of $v,~B_{(s,t)},~c_{(s,t)}$ and $\epsilon_s$
in terms of $q$ (defined earlier), corresponding to scalar and power amplitude. To get insight, we have plotted
$A_s-q$ in the left panel of Fig. \textbf{5} which shows that for all values of $m>1$, the value of scalar amplitude
is always less than unity. Now, we are able to evaluate the energy density using Eqs.(\ref{32}), (\ref{32a}) and first field equation.
As we know that at the earliest stages of the cosmic evolution (not long after the singularity), $H_2$ and
$\rho$ might have been arbitrarily large. It is usually assumed that at densities $\rho\gtrsim M^{4}_P\sim10^{94}g/cm^3$,
quantum gravity effects are so significant that quantum fluctuations of the metric exceed the classical value of
$g_{\mu\nu}$, and classical space-time does not provide an adequate description of the universe. Hence, to
prove that our model does not lie in quantum gravity regime, we have plotted $\rho$ versus $q$ in the right panel of
Fig.\textbf{5} for three different values of $m$. It is clear that there is an increasing behavior among $\rho,~q$ and $m$.
An upper bound for anisotropic parameter $1<m<45$ is also calculated for which $\rho\lesssim M^{4}_P\sim35.1557\times10^{72}(GeV)^4$
in the range $q>0$.

These quantities lead us to calculate an important physical parameter,
i.e., tensor-scalar spectrum ratio, given by
\begin{equation}\nonumber
r=\frac{P_t}{P_s}=\left(\frac{A_t}{A_s}\right)^2=\frac{q(m+5)+6(m+2)}
{(1-(\frac{1+2m}{m+2})(\frac{q}{q+2}))(q(5m+7)+6(m+2))}.
\end{equation}
The behavior of tensor-scalar ratio with respect to $q$ is checked in the left graph of
Fig.\textbf{6}. The Planck data put an upper bound on the physical parameter $r$, i.e.,
$r<0.11~ (95\%$ C.L.). The graphical analysis (shown in the left graph of Fig.\textbf{6})
proves that the considered anisotropic model is compatible with recent astrophysical data
presented by Planck collaboration for all values of $m\in(1,\infty)$.
Moreover, the best fit value of $r$ is obtained in the interval $q\in(-3,-2)$. The right plot of
Fig.\textbf{6} shows that anisotropic model is not viable in the stable range $0<q<1$ as the $r-q$ trajectory 
goes away from the standard value.

\begin{figure}
\centering\epsfig{file=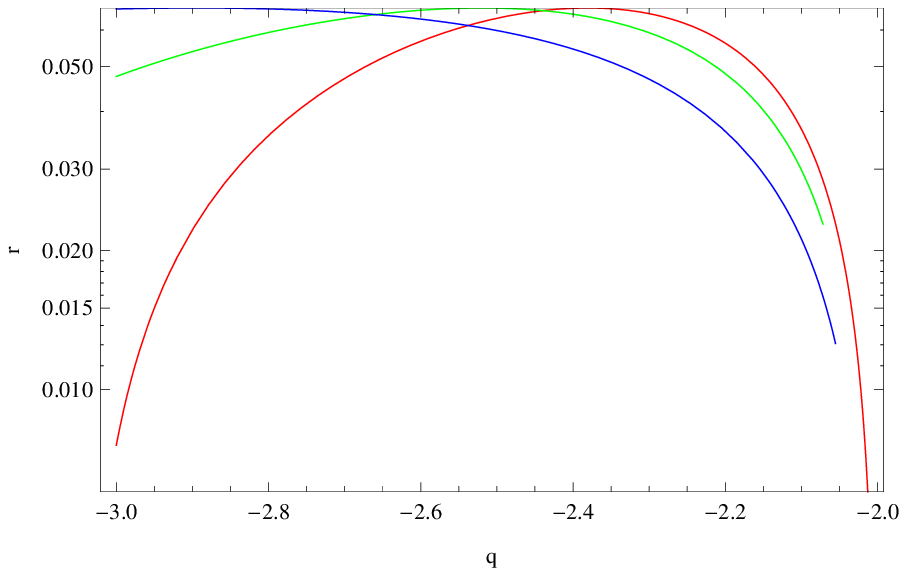,
width=0.55\linewidth}\epsfig{file=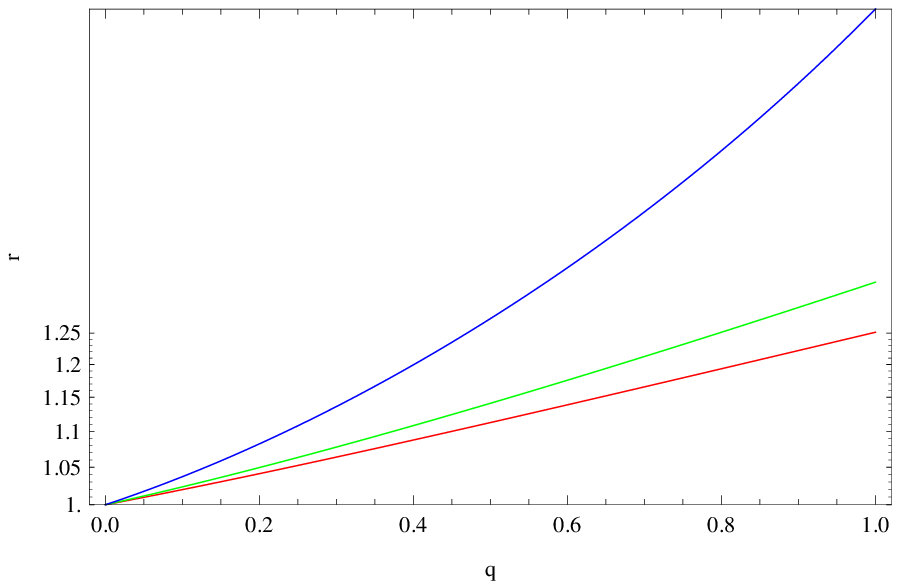,
width=0.55\linewidth}\caption{(left) $r$ versus $q$: for
$m=1.1$ (red); $m=2.5$ (green); $m=10^3$ (blue), (right) $r$ versus $q$ for $q>0$.}
\end{figure}
\begin{figure}
\center\epsfig{file=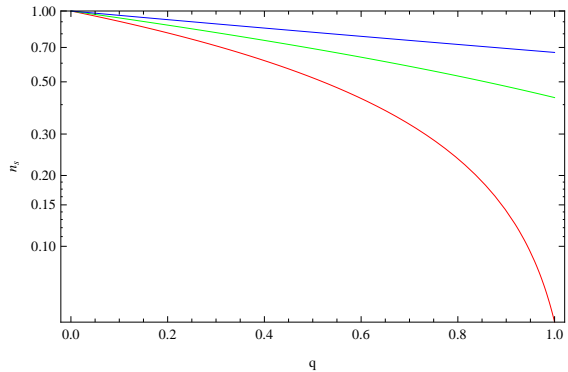,
width=0.55\linewidth}\caption{$n_s$ versus $q$: for
$m=1.1$ (red); $m=2.5$ (green); $m=45$ (blue).}
\end{figure}
Now, we are able to find another parameter, the spectral index $(n_s)$ by the following formula
\begin{equation}\nonumber
n_s-1=\frac{d\ln P_s}{d\ln k}\mid_{c_sk=bH_2}=\frac{d\ln P_s}{dt}\frac{dt}{d\ln k}\mid_{c_sk=bH_2},
\end{equation}
where
\begin{equation}\nonumber
\frac{d\ln k}{dt}=\left(\frac{m+2}{3}\right)H_2\left[1-\left(\frac{3}{m+2}\right)\epsilon\right].
\end{equation}
Putting in the above equation, we get $n_s$ in terms of $q$ as follows
\begin{equation}\nonumber
n_s-1=-\frac{6\epsilon}{(m+2)-3\epsilon}=-\frac{6q}{(m+2)(q+2)-3q}.
\end{equation}
Figure \textbf{7} proves the compatibility of our anisotropic model with recent Planck
astrophysical data, i.e, $n_s=0.9608\pm0.0054 ~(68\%$C.L. or $1\sigma$ error). I have plotted
$n_s$ versus $q$ for three different values of the anisotropic parameter $m=1.1,~2.5,~45$, picked from the
range of $1<m<\infty$. It is clear from Fig. \textbf{7} that for all these values of $m$,
$n_s$ lies in the range $0<n_s<1$. It is also noticed that for $m>45$, the value of $n_s$ exceeds from unity
which is not a physical value. Hence in this case, the model is compatible with Planck data for $1<m<45$.

\section{Concluding Remarks}

Recent astrophysical data coming from the Planck satellite verifies that the
``large angle anomalies" represent original feature of the cosmic CMB map.
This result play a vital role to consider that the ``small temperature anisotropies"
and ``large angle anomalies" may be influenced by anisotropic phase during the
early cosmic evolution. This statement has a key importance as it favors to
develop an alternative cosmic model to interpret the effects of the early-time
universe on the current structure of large scale without affecting the processes of
nucleosynthesis. Warm inflation is a good model for LSS formation, in which
the density fluctuations arise from thermal fluctuation. It is described by a damping factor
in the inflaton's equation of motion. The magnitude of this factor suggests the
prospect that it has the strong effect prolonging inflation.

Motivated by this fact, I study the warm inflation with non-minimal derivative coupling model
(which solves the issue of few number of e-folds and clear the scenario ``how reheating occurs?")
during rapid oscillations. To get the comprehensive results, I have used the framework of homogeneous but
anisotropic LRS BI cosmic model, which is asymptotically equivalent to the standard FRW universe. I
reconstruct the formalism of oscillatory inflation using anisotropic background. A power law form of potential
is used to find the adiabatic index $(\gamma)$ and time period of the oscillations. On solving these equations,
I am able to calculate a constraint on the amplitude of the inflaton $(\Phi)$ during non-minimal case for the
realization of inflation. To check the validity of the inequality in anisotropic model, I have plotted trajectories
of $\Phi^{\frac{q+2}{2}}$ and the expression on right hand side (say $f$) of Eq.(\ref{12}) versus $q$ for specified
values of $m$ in the left and right panel of Fig.\textbf{1}, respectively. It is very much clear from the comparison
of left and right graph of Fig.\textbf{1} that the value of $\Phi^{\frac{q+2}{2}}$ is much less than the expression
$f$ for $q>0$ and $m>0;~m\neq1$ (the result also holds for $q<0$).

The end of inflation is discussed by using a special form of Damour-Mukhanov potential where
the inflation continues whenever $d<g(\mathfrak{b},q)$ (given in Bezrukov 2008). The violation of
this inequality (i.e., $d=g(\mathfrak{b}_{end},q)$ and $d>g(\mathfrak{b}(t>t_{end}),q)$),
inflation terminates. The graphical analysis of this double valued function is shown in
Fig. \textbf{2} for three different values of $\mathfrak{b}=1$ (red),~$5$ (green)
~$10$ (blue). It is observed that $g(\mathfrak{b},q)$ has an increasing behavior for all
values of $\mathfrak{b}>0$ in the range $q\in(-\infty,1)$. The point of intersection of
the three curves is $(1,0)$, so inflation ends for $d>1$ and $d\simeq1~ (\phi\sim\phi_c)$.
Further, I calculate the number of e-folds for the minimal and non-minimal case during high
friction regime. The plot for $\mathcal{N}-q$ is presented to set an upper bound on $\mathcal{N}_{min}$,
by fixing $\Phi_{end}\sim10^{-17}m_P$ (left panel) and $\Phi_{end}\sim10^{-6}m_P$ (right panel)
and varying $m=1.5,~2.5,~10$ in Fig.\textbf{3}. It is noticed from both graphs of Fig.\textbf{3}
that an increment in the scale of $\Phi_{end}$ leads to decrease the value of $\mathcal{N}_{min}$
while an increasing relationship exists between $\mathcal{N}$ and $m$. The theory may become more
viable at least in the context of perturbations generation as the anisotropic model has ability to
provide more number of e-folds ($\mathcal{N}>8.4$) in non-minimal case as compared to the minimal case.

Moreover, the cosmological perturbation scheme is developed in the anisotropic background
using the Mukhanov-Sasaki equation during rapid oscillation era. The explicit expressions
are calculated for speed of sound, scalar and tensor power spectrum, tensor-scalar spectrum ratio
and spectral index. Figure \textbf{4} shows that the speed of sound lies in the feasible range
$0<c_s<1$ for all values of $m>0;~m\neq1$. To get insight, we have plotted $A_s-q$ in the left panel of
Fig. \textbf{5} which shows that for all values of $m>1$, the value of scalar amplitude
is always less than unity. Hence, to prove that our model does not lie in quantum gravity regime,
we have plotted $\rho$ versus $q$ in the right panel of Fig.\textbf{5} for three different values of $m$.
It is clear that there is an increasing behavior among $\rho,~q$ and $m$. An upper bound for anisotropic
parameter $1<m<45$ is also calculated for which $\rho\lesssim M^{4}_P\sim35.1557\times10^{72}(GeV)^4$
in the range $q>0$. The behavior of tensor-scalar ratio with respect to $q$ is checked in the left graph of
Fig.\textbf{6}. The left graph of Fig.\textbf{6} proves that the considered anisotropic model is compatible with
recent astrophysical data presented by Planck collaboration for all values of $m\in(1,\infty)$.
Moreover, the best fit value of $r$ is obtained in the interval $q\in(-3,-2)$. While the parameter $r$ remains 
incompatible with recent data for a stable range of $q>0$ (right plot of Fig.\textbf{6}). $n_s$ versus $q$ is plotted
for three different values of the anisotropic parameter $m=1.1,~2.5,~45$, picked from the
range of $1<m<\infty$ in Fig.\textbf{7}. It is clear from graphical analysis that for all these values of $m$,
$n_s$ lies in the range $0<n_s<1$. It is also noticed that for $m>45$, the value of $n_s$ exceeds from unity
which is not a physical value. Hence in this case, the model is compatible with Planck data for $1<m<45$.

It is worth mentioned that all the results reduced to the isotropic case for $m=1$ (Sadjadi and Goodarzi 2014). 
The major difference is that it is not possible to find the exact value of the parameter $q$ during anisotropic 
background. It is noticed that the stable models can not be achieved for $q<0$ as compared to Cembranos et al. 
(2016). In future, I will perform this type of study by considering thermal correction to the effective potential, 
also the temperature dependency of the dissipative factor and checking all consistency conditions.

\end{document}